\input harvmac 

\input epsf
\ifx\epsfbox\UnDeFiNeD\message{(NO epsf.tex, FIGURES WILL BE
IGNORED)}
\def\figin#1{\vskip2in}
\else\message{(FIGURES WILL BE INCLUDED)}\def\figin#1{#1}\fi
\def\ifig#1#2#3{\xdef#1{fig.~\the\figno}
\goodbreak\topinsert\figin{\centerline{#3}}%
\smallskip\centerline{\vbox{\baselineskip12pt
\advance\hsize by -1truein\noindent{\bf Fig.~\the\figno:} #2}}
\bigskip\endinsert\global\advance\figno by1}


\overfullrule=0pt
\def\Title#1#2{\rightline{#1}\ifx\answ\bigans\nopagenumbers\pageno0\vskip1in
\else\pageno1\vskip.8in\fi \centerline{\titlefont #2}\vskip .5in}

scaled\magstep3
  
scaled\magstep3
  
scaled\magstep3
  
scaled\magstep3

\lref\mal{J.~Maldacena, 
``The large N limit of superconformal field theories and 
supergravity,'' 
Adv.\ Theor.\ Math.\ Phys.\ {\bf 2}, 231 (1998) [Int.\ J.\ Theor.\ 
Phys.\ {\bf 38}, 1113 (1998)] [hep-th/9711200].}

\lref\bupepol{A.~Buchel, A.~W.~Peet and J.~Polchinski,
``Gauge dual and noncommutative extension of an N = 2 supergravity  
solution,''
Phys.\ Rev.\ D {\bf 63}, 044009 (2001)
[hep-th/0008076].}

\lref\jppol{C.~V.~Johnson, A.~W.~Peet and J.~Polchinski,
``Gauge theory and the excision of repulson singularities,''
Phys.\ Rev.\ D {\bf 61}, 086001 (2000)
[hep-th/9911161].}

\lref\jonlopage{C.~V.~Johnson, K.~J.~Lovis and D.~C.~Page,
``The Kaehler structure of supersymmetric holographic RG flows,''
[hep-th/0107261].}

\lref\malis{N.~Itzhaki, J.~M.~Maldacena, J.~Sonnenschein and 
S.~Yankielowicz,
``Supergravity and the large N limit of theories with sixteen  
supercharges,''
Phys.\ Rev.\ D {\bf 58}, 046004 (1998)
[hep-th/9802042].}

\lref\malnun{J. Maldacena and C. N\'u\~nez, ``Supergravity description
of field theories on curved manifolds and a no go theorem'',
Int.J.Mod.Phys. {\bf A16} (2001) 822, [hep-th/0008001].} 

\lref\malnu{J.~M.~Maldacena and C.~N\'u\~nez,
``Towards the large N limit of pure N = 1 super Yang Mills,''
Phys.\ Rev.\ Lett.\  {\bf 86}, 588 (2001)
[hep-th/0008001].}

\lref\chp{M. Cvetic, H. Lu and C.N. Pope, ``Massless 3-brane in
M-theory'', [hep-th/0105096].}%

\lref\mnas{
J.~Maldacena and H.~Nastase, 
``The supergravity dual of a theory with dynamical supersymmetry
breaking,'' 
[hep-th/0105049]. }

\lref\jerome {
J.~P.~Gauntlett, N.~Kim, D.~Martelli and D.~Waldram, 
``Wrapped fivebranes and N = 2 super Yang-Mills theory,'' 
[hep-th/0106117]. }

\lref\zaff{F.~Bigazzi, A.~L.~Cotrone and A.~Zaffaroni, 
``N = 2 gauge theories from wrapped five-branes,'' [hep-th/0106160].} 

\lref\oht{K.~Ohta, 
``Moduli space of vacua of supersymmetric Chern-Simons theories and 
type IIB branes,'' JHEP {\bf 9906}, 025 (1999) [hep-th/9904118]. }

\lref\ohtm{B.~Lee, H.~Lee, N.~Ohta and H.~S.~Yang, 
``Maxwell Chern-Simons solitons from type IIB string theory,'' Phys.\ 
Rev.\ D {\bf 60}, 
106003 (1999) [hep-th/9904181].} 

\lref\berg{O.~Bergman, A.~Hanany, A.~Karch and B.~Kol, 
``Branes and supersymmetry breaking in 3D gauge theories,'' JHEP {\bf 
9910}, 036 (1999) [hep-th/9908075]. }

\lref\gmat{
J.~Gomis and T.~Mateos,
``D6 branes wrapping Kaehler four-cycles,''
[hep-th/0108080].}

\lref\clpo{ M. Cvetic, H. Lu  and C.N. Pope, ``Consistent Kaluza-Klein
Sphere Reductions'', Phys.Rev. {\bf D62} (2000) 064028.}%

\lref\bsv{ M. Bershadsky, V. Sadov and  C. Vafa, ``D-Branes and
  Topological Field Theories'', Nucl.Phys. {\bf B463} (1996) 420.}%

\lref\gomis{
J.~Gomis,
``D-branes, holonomy and M-theory,''
Nucl.\ Phys.\ B {\bf 606}, 3 (2001)
[hep-th/0103115].}

\lref\ejp{
N.~Evans, C.~V.~Johnson and M.~Petrini,
``The enhancon and N = 2 gauge theory/gravity RG flows,''
JHEP {\bf 0010}, 022 (2000)
[hep-th/0008081].}

\lref\jpp{
C.~V.~Johnson, A.~W.~Peet and J.~Polchinski,
``Gauge theory and the excision of repulson singularities,''
Phys.\ Rev.\ D {\bf 61}, 086001 (2000)
[hep-th/9911161].}

\lref\gkw{
J.~P.~Gauntlett, N.~Kim and D.~Waldram,
``M-fivebranes wrapped on supersymmetric cycles,''
Phys.\ Rev.\ D {\bf 63}, 126001 (2001)
[hep-th/0012195].}

\lref\acha{B.~S.~Acharya, J.~P.~Gauntlett and N.~Kim,
``Fivebranes wrapped on associative three-cycles,''
Phys.\ Rev.\ D {\bf 63}, 106003 (2001)
[hep-th/0011190].}

\lref\witten {Witten, ``Supersymmetric index...", 9903005.}

\lref\gkpw{ Jerome P. Gauntlett, N. Kim, S. Pakis and D. Waldram,
``Membranes Wrapped on Holomorphic Curves'', [hep-th/0105250].}%

\lref\noz{
H.~Nieder and Y.~Oz,
``Supergravity and D-branes wrapping special Lagrangian cycles,''
JHEP {\bf 0103}, 008 (2001)
[hep-th/0011288].}

\lref\edel{J.~D.~Edelstein and C.~N\' u\~ nez,
``D6 branes and M-theory geometrical transitions from gauged  
supergravity,''
JHEP {\bf 0104}, 028 (2001)
[hep-th/0103167].}
\lref\hern{R. Hern\'andez, ``Branes Wrapped on Coassociative Cycles'',
hep-th/0106055.}%

\lref\schv{
M.~Schvellinger and T.~A.~Tran,
supergravity in five dimensions,''
JHEP {\bf 0106}, 025 (2001)
[hep-th/0105019].}

\lref\jerkim{J.P. Gauntlett and N. Kim, ``M-Fivebranes Wrapped on
Supersymmetric Cycles II'', [hep-th/0109039].}%

\lref\npst{C. N\'u\~nez, I.Y. Park, M. Schvellinger and T.A. Tran,
`` Supergravity duals of gauge theories from F(4) gauged supergravity
in six dimensions'', JHEP {\bf 0104} (2001) 025.}%

\lref\ahwit{I. Affleck, J. Harvey and E. Witten'', ``Instantons and
(Super-)symmetry breaking in (2+1) Dimensions, Nucl. Phys. {\bf 206}
(1982) 413.}%

\lref\PILCH{
K.~Pilch and N.~P.~Warner,
``N = 2 supersymmetric RG flows and the IIB dilaton,''
Nucl.\ Phys.\ B {\bf 594}, 209 (2001)
[hep-th/0004063].}

\lref\gubser{S.~S.~Gubser,
``Curvature singularities: The good, the bad, and the naked,''
hep-th/0002160.}

\lref\cpwar{
R.~Corrado, K.~Pilch and N.~P.~Warner,
``An N = 2 supersymmetric membrane flow,''
hep-th/0107220.}

\lref\khawar{A.~Khavaev and N.~P.~Warner,
``An N = 1 supersymmetric Coulomb flow in IIB supergravity,''
hep-th/0106032.}

\lref\jonlopageb{C.~V.~Johnson, K.~J.~Lovis and D.~C.~Page,
``Probing some N = 1 AdS/CFT RG flows,''
JHEP {\bf 0105}, 036 (2001)
[hep-th/0011166].}


\lref\ppvn{
M.~Pernici, K.~Pilch and P.~van Nieuwenhuizen,
Phys.\ Lett.\ B {\bf 143}, 103 (1984).}


\Title{\vbox{\baselineskip12pt \hbox{hep-th/0109177}
\hbox{CALT-68-2350}
\hbox{CITUSC/01-033}}}
{\vbox{
\centerline {$D=2+1$ ${\cal N}=2$ Yang-Mills Theory From Wrapped 
Branes}
}}
%
%
\centerline{Jaume Gomis ~$^{a}$\ and \ Jorge G. Russo~$^{b}$ }

\bigskip
\centerline{$^{a}$ {\it Department of Physics,
California Institute of Technology} }
\centerline{\it Pasadena, CA 91125}

\medskip
\centerline{$^{b}$ {\it Departamento de F\'\i sica,
Universidad de Buenos Aires} }
\centerline{\it Ciudad Universitaria, 1428 Buenos Aires}

%
%

\def\[{\left [}
\def\]{\right ]}
\def\({\left (}
\def\){\right )}
\def\td{\tilde }
\def\p{\partial}
\def\a{\alpha }
\def\b{\beta }
\def\ha{ {\textstyle{1\over 2}}}
\def\N{ {\cal N} }

\vskip .3in

\centerline{\bf Abstract}

We find a new solution of Type IIB supergravity which represents a 
collection 
of D5 branes wrapped on the topologically non-trivial  
${\bf S}^3$ of the deformed conifold geometry $T^*{\bf S}^3$. The Type
IIB 
solution is obtained by lifting a new solution of $D=7$ $SU(2)_L\times
SU(2)_R$ gauged supergravity to ten dimensions in which $SU(2)_D$
gauge fields in the 
diagonal subgroup are turned on. The supergravity solution describes
a slice of the Coulomb branch in
the large N limit of ${\cal N}=2$ SYM in three dimensions.

\smallskip
\noindent
\Date{}


\newsec{Introduction}

Supersymmetric gauge theories can be realized as the low
energy effective field theory living on a collection of branes wrapping
supersymmetric cycles \malnun . The topological twisting which is 
required  to
make these gauge theories supersymmetric \bsv\ can be incorporated
in the supergravity realization of these wrapped branes by turning on
non-trivial gauge fields in gauged supergravity. These gauged
supergravity solutions can then be lifted to ten or eleven dimensions
and give rise a host of interesting dual descriptions to the infrared 
dynamics of 
various gauge theories 
\refs{\malnu\acha\noz\gkw\npst\edel\schv\mnas\chp\gkpw\hern
\jerome\zaff\gmat -\jerkim }.

In this paper we find the supergravity solution representing a
collection of $N$ NS5-branes wrapping the supersymmetric ${\bf S}^3$ of
the deformed conifold geometry, which admits a metric with $SU(3)$
holonomy. This supergravity solution describes the infrared dynamics
of three dimensional ${\cal N}=2$ $SU(N)$ Super-Yang-Mills (SYM). 

We find this gravity dual by finding a solution of
$SO(4)=SU(2)_L\times SU(2)_R$ gauged
supergravity. Supersymmetry requires that we turn on $SU(2)_D$ gauge
fields of gauged supergravity.
Our solution provides an example of
an ansatz with $SU(2)_D=(SU(2)_L\times SU(2)_R)_{diag}$ gauge fields 
that solves the full $SO(4)$ gauged
supergravity equations of motion,
despite the fact that an $SU(2)_D$ truncation of $SO(4)$ gauged
supergravity is in general 
inconsistent. This solution can then be lifted to
ten dimensions and describes certain aspects of the gauge theory 
dynamics.

Previous constructions of wrapped branes representing
three dimensional SYM appeared \refs{\acha, \gomis, \mnas, \gmat}(see
also \cpwar ).
SYM theory in three dimensions
can include, in addition to the
 usual YM term,  
 a Chern-Simons term in the action with level $k$. In the
 supergravity description based on  D5 branes 
 wrapping  ${\bf S}^3$, this term arises when the background has a
 non-trivial RR field $H_3^{R}$ with components 
 in the ${\bf S}^3 $ directions  \mnas .
The present supergravity solutions have $H_3^{R}=0$  
in the ${\bf S}^3 $ directions, so they describe 2+1 SYM without 
Chern-Simons
term ($k=0)$.

The solutions describe a slice of the Coulomb branch of the gauge
theory, which is 
parametrized by the $N-1$ real scalars of the vector multiplet
together with the $N-1$ scalars obtained upon dualization of the
photons\foot{The real scalar of the vector multiplet arises from
dimensionally reducing the four dimensional ${\cal N}=1$ vector
multiplet.}. There is a 
curvature singularity in the supergravity solution, related to 
the NS5-brane sources. We analyze the moduli space of the gauge theory
by sending a probe into the geometry and reproduce within supergravity
the existence of a complex moduli space with a Kahler metric, as
required by supersymmetry.

It is well known that monopoles in this theory generate a 
nonperturbative
superpotential which destabilizes the Coulomb branch \refs{\ahwit}. 
On the other
hand, our supergravity solutions exhibits a Coulomb branch. This is
not inconsistent, since the effects that induce lifting of the Coulomb
branch are of order $e^{-N}$ in the t'Hooft limit, so 
they are not visible in the supergravity approximation.

This paper is organized as follows. In section $2$ we
describe the low energy effective field theory on the wrapped branes
and study the conditions under which the required $SU(2)_D$ truncation
can give a solution. In section $3$ we find first order equations
which solve the seven dimensional gauged supergravity equations of
motion. In section $4$  
the solution is lifted to ten dimensions, giving a one-parameter family 
of solutions of type II
supergravity. We discuss the behavior of the solutions for different 
values of the parameter.
 Section $5$ makes
contact with field theory expectations. We calculate the effective
action on the Coulomb branch of the gauge theory in the large $N$
limit. Finally, in the appendix, we discuss an ansatz for NS5 branes 
with
$SU(2)_D$ gauge fields, with the aim of describing four-dimensional 
$\N=2 $ SYM.

\newsec{Wrapped Brane Realization of Three Dimensional {\cal N}=2
$SU(N)$ SYM}

This gauge theory can be realized by wrapping $N$ NS5-branes on the 
${\bf
  S}^3$  of the deformed conifold geometry $T^* {\bf S}^3$, which
  admits a metric with $SU(3)$ holonomy.
In order to obtain a globally
supersymmetric gauge theory one must couple the field theory on the
 branes to the gauge fields which couple to the R-symmetry currents of 
five-branes in flat space, that is, the field theory must be
  topologically twisted
\bsv . Then, supersymmetry can be realized by
finding  
 constant  and  
``covariantly"
constant spinors on ${\bf S}^3$ \foot{The standard Dirac equation on
 ${\bf S}^3$ does 
not admit zero modes and therefore an untwisted  supersymmetric theory
 cannot  be
realized on it.}
\eqn\gen{\eqalign{
\partial \epsilon&=0\ ,
\cr
{\cal D}\epsilon &=\left(\partial +{1\over
    4}\omega_{ab}\Gamma^{ab}+{1\over
    4}A_{ij}\gamma^{ij}\right)\epsilon=0\ ,
}}
where $a,b$ are tangent space indices on the ${\bf S}^3$ and
$i,j$ are vector indices of the $SO(4)=SU(2)_L\times SU(2)_R$
    $R$-symmetry group of the 
five-branes in flat space. 

Turning this background gauge field effectively changes the
transformation properties of all fields under a modified tangent space
symmetry group of ${\bf S}^3$,  under which the unbroken
supersymmetries transform as  scalars. The topological
twisting giving rise
 to a three dimensional ${\cal N}=2$ theory 
is obtained by  identifying the $SU(2)_{S^3}$ spin
  connection with the $SU(2)_D=(SU(2)_L\times SU(2)_R)_{\rm diag}$
subgroup of the $SU(2)_L\times SU(2)_R$ R-symmetry group of the 
NS5-brane so that the
supersymmetry generators are singlets under $SU(2)_{\rm
diag}=(SU(2)_{S^3}\times 
SU(2)_D)_{\rm diag}$.\foot{A different twisting, used previously in
\malnu , in which one identifies the $SU(2)_{S^3}$ spin
  connection with the $SU(2)_L$ subgroup of the $SU(2)_L\times
SU(2)_R$ 
R-symmetry group, gives rise to a four-dimensional  ${\cal N}=1$
theory. In that case  the
supersymmetry generators are singlets under $SU(2)_{\rm
diag}=(SU(2)_{S^3}\times
SU(2)_L)_{\rm diag}$.}

To verify that the field content that follows from this twist gives
rise to a three dimensional ${\cal N}=2$  massless vector multiplet,
one decomposes the original six dimensional ${\cal N}=(1,1)$
vector multiplet on the five-branes under the symmetries left unbroken
by the background, which are  $SO(1,2)\times SU(2)_{S^3}\times 
SU(2)_L\times SU(2)_R$
\eqn\spcca{\eqalign{
\hbox{gauge}:&\ ({\bf 3},{\bf 1},{\bf 1},{\bf 1})\oplus ({\bf 1},{\bf
3},{\bf 1},{\bf 1}) \cr
\hbox{scalars}:&\ ({\bf 1},{\bf 1},{\bf 2},{\bf 2}) \cr
\hbox{spinors}:&\ ({\bf 2},{\bf 2},{\bf 2},{\bf 1})\oplus ({\bf 2},{\bf
2},{\bf 1},{\bf 2}). \cr}}
Then it follows that the transformation properties of these modes
under $SO(1,2)\times SU(2)_{\rm{diag}}$ are
\eqn\spccab{\eqalign{
\hbox{gauge}:&\ ({\bf 3},{\bf 1})\oplus ({\bf 1},{\bf
3}) \cr
\hbox{scalars}:&\ ({\bf 1},{\bf 1}) \oplus ({\bf 1},{\bf 3})\cr
\hbox{spinors}:&\ 2({\bf 2},{\bf 1})\oplus 2({\bf 2},{\bf
3}). \cr}}
Therefore, the modes singlet under $SU(2)_{\rm{diag}}$ fill the ${\cal
N}=2$ vector multiplet.

In supergravity, the $SO(4)$ global $R$-symmetry group of a
five-brane appears 
as a gauge symmetry of the $7$-dimensional effective supergravity
theory obtained by 
reducing Type II supergravity on the transverse ${\widetilde{\bf 
S}}^{3}$ which
appears in the near horizon solution of the five-brane.\foot{
We recall that the near horizon geometry of a five-brane is
${\bf R}^{1,5}\times{\bf R}\times {\widetilde{\bf S}}^3$ with a linear 
dilaton
along ${\bf R}$.} 
Therefore, $7$-dimensional gauged supergravity is a natural arena  to 
construct
supergravity 
solutions in which a background R-symmetry gauge field is turned on.
solution

The bosonic content of the $SO(4)$ gauged supergravity  is the
graviton, a two-form, $SO(4)$ gauge fields $A_{ij}$ and ten scalars
$T_{ij}$ transforming in the two-index symmetric representation of
$SO(4)=SU(2)_L\times SU(2)_R$
\eqn\transform{\eqalign{
\hbox{gauge fields}\ A_{ij}&: ({\bf 3},{\bf 1})\oplus ({\bf 1},{\bf 
3})\cr
\hbox{scalars}\ T_{ij} &: ({\bf 1},{\bf 1})\oplus({\bf 3},{\bf 3})\ .
}}
By turning on some fields in this gauged supergravity,
one can  construct solutions having a space-time interpretation as
wrapped branes.

The bosonic Lagrangian of  $SO(4)$ gauged supergravity in the absence
of the two-form is given by \ppvn 
\eqn\Lagrangian{\eqalign{
{\cal L}=\sqrt{g}\Big(&R-{5\over
16}Y^{-2}\partial_{\mu}Y\partial^{\mu}Y-{1\over
4}\widetilde{T}_{ij}^{-1}\widetilde{T}_{kl}^{-1}D_\mu\widetilde{T}_{jk}
D^\mu\widetilde{T}_{li}-
{1\over
8}Y^{-1/2}\widetilde{T}_{ik}^{-1}
\widetilde{T}_{jl}^{-1}F^{ij}_{\mu\nu}F_{kl}^{\mu\nu}\cr  
&-{1\over 2}g^2 Y^{1/2}(2\widetilde{T}_{ij}\widetilde{T}_{ij}-
(\widetilde{T}_{ii})^2)\Big),}}
where 
\eqn\defini{\eqalign{
Y&=\hbox{det}(T_{ij})\ ,\cr
\widetilde{T}_{ij}&=Y^{-1/4}T_{ij}\ ,\cr
DT_{ij}&=dT_{ij}+g[A,T]_{ij}\ .
}}
Any solution to the $SO(4)$ gauged supergravity equations lifts to a
solution of ten dimensional Type IIB supergravity.  
Explicit  formulas  \clpo\ for the ten dimensional fields in terms
of the seven dimensional ones are given in  section 4.

For the case under study, one needs to turn on $SU(2)_D$ gauge 
fields $A_a$. These can be identified as the following components of
the $SO(4)$ gauge fields $A_{ij}$
\eqn\gaugefi{
A_1\equiv A_{23}\qquad A_2\equiv -A_{13}\qquad A_3\equiv A_{12}.
}
The $SU(2)_D$ gauge fields may in general act as
non-linear
sources for the gauge fields which have been set to zero.
The reason is that they are sources for the fields 
$A_{14},A_{24},A_{34}$.
We have studied the $SO(4)$ gauged supergravity equations following
from \Lagrangian\ and found  
sufficient conditions that have to be
met in order for a solution with only $SU(2)_D$ gauge fields to be 
compatible 
with the full $SO(4)$ equations of motion. They can be summarized as 
follows: 

\smallskip
\noindent $\bullet$ Only $SO(4)$ matter modes of gauged supergravity
that are singlets 
under $SU(2)_D$ can be turned on. Since the non-trivial $SO(4)$ gauged
supergravity modes transform under $SU(2)_D$ as 
\eqn\trunct{\eqalign{
\hbox{gauge fields}&: 2\;{\bf 3}\cr
\hbox{scalars}&: 2\; {\bf 1}\oplus {\bf 3}\oplus {\bf 5},}}
then only $SU(2)_D$ gauge fields $A_a$, 2 real scalars, the two-form 
and the
metric can be turned on. Moreover, invariance under $SU(2)_D$ requires
the scalar matrix to take the following form\foot{This can be easily
understood as follows. The $SO(4)$ generators $J_{ij}=-J_{ji}$ can be 
divided
into boost generators $K_a$ and rotation generators $J_a$. The 
$SU(2)_D$
subgroup of $SO(4)$ is generated by $J_a$, and the most general
four-dimensional symmetric matrix invariant under rotations is
of the form (2.9).
}
\eqn\scal{
T_{ij}=e^{y/4}\hbox{diag}(e^x,e^x,e^x,e^{-3x})\ .
}

\noindent 
$\bullet$ The $SU(2)_D$ gauge fields must satisfy the following 
condition 
\eqn\consistent{
F_a\wedge *F_b=0\ \hbox{for}\ a\neq  b.
}

Once these conditions are satisfied one can perform an $SU(2)_D$
reduction of the $SO(4)$ gauged supergravity Lagrangian 
\Lagrangian , giving the     
 $SU(2)_D$ Lagrangian 
\eqn\truncLagr{\eqalign{
{\cal L}=\sqrt{g}\Big(&R-{5\over 16}\partial_{\mu}y\partial^{\mu}y-3
\partial_{\mu}x\partial^{\mu}x-{1\over
4}e^{-2x-y/2}(F_{\mu\nu}^1F^{\mu\nu}_1+F_{\mu\nu}^2F^{\mu\nu}_2+
F_{\mu\nu}^3F^{\mu\nu}_3)\cr
&+{1\over
2}g^2e^{y/2}(3e^{2x}+6e^{-2x}-e^{-6x})\Big).
}}
Any solution of this Lagrangian with gauge fields satisfying 
\consistent\
fully solves 
the $SO(4)$ gauged supergravity equations, and it can be lifted as
a solution of ten dimensional Type IIB
supergravity.

\newsec{The Supergravity Solution in $D=7$}

In order to construct the ${\cal N}=2$ supersymmetric solution, we
 will consider   
 seven dimensional  $SO(4)$ gauged supergravity, in which only 
$SU(2)_D$ gauge
 fields are turned on. The worldvolume of the five-branes we are
 interested in is ${\bf R}^{1,2}\times {\bf S}^3$.
The metric ansatz compatible with the worldvolume of the five-branes  
is
\eqn\metr{
ds_7^2=e^{2f(r)}(ds^2({\bf R^{1,2}})+dr^2)+{a^2(r)\over 4}w_aw_a \ ,
}
where $w_a$ are left invariant  $SU(2)$ one-forms satisfying
\eqn\satis{
dw_{a}={1\over 2}\epsilon_{abc}w_b\wedge w_{c}\ .
} 
One can describe them in Euler angle parametrization,
$$
w_1=\cos\b d\theta + \sin \b \sin\theta d\varphi\ ,\ \ \ 
w_2=\sin\b d\theta - \cos \b \sin\theta d\varphi\ ,\ \  \ 
w_3=d\b +\cos\theta  d\varphi \ .
$$
where the angles $\beta,\theta$ and $\varphi$ parametrize the ${\bf
S}^3$ which the branes are wrapping.

The identification of the $SU(2)_{S^3}$ spin
  connection with the $SU(2)_D$ gauge fields that follows from \gen\
  requires that  
\eqn\gauge{
A_a=-{1\over 2g}w_a\ ,
}
with the $A_a$'s given in \gaugefi , and the corresponding curvature is 
given by
\eqn\curv{ 
F_a=-{1\over 8g}\epsilon_{abc}w_b\wedge w_c\ .
}
We note that this gauge field configuration is compatible with the 
$SO(4)$ gauged
supergravity equations since $F^a\wedge * F^b=0$ for $a\neq b$. 

The string metric for wrapped NS5-branes typically has no warp factor 
in 
the parallel, flat directions along the brane.
 Since the ten dimensional string frame metric is of the form \clpo ,
$ds^2_{st}=e^{2f+y/2}ds^2({\bf R^{1,2}})+...$ (see below in section
$5$), it is natural to look for solutions with  $y(r)=-4f(r)$. It is
easy to show that this ansatz   
is compatible with  the equations of motion that follow from
\truncLagr .

We find the equations of motion by reducing the problem to that of an
effective quantum mechanics for the ansatz, which together with a
Hamiltonian constraint results in the gauged supergravity equations of
motion. Using the ansatz \metr , \gauge\ together  with $y=-4f$ 
the various terms in the action \truncLagr\ take the following form
\eqn\pluffedin{\eqalign{
{\cal L}_R&=e^{2f}a^3\left(6{e^{2f(r)}\over a^2}+6{\left({\dot
a}\over a\right)^2}+18{{\dot
a}\over a}{\dot f}+6 {\dot f}^2\right)\ , 
\cr
{\cal L}_{scal}&=-e^{2f}a^3\left({5}{\dot f}^2+3{\dot
x}^2\right)\ ,
\cr
{\cal L}_{gauge}&=-{1\over 2g^2}e^{2f}a^3 \left( {e^{4f-2x}\over 
a^4}\right)\ ,
\cr
{\cal L}_{pot}&={1\over 2}e^{2f}a^3 g^2(3e^{2x}+6e^{-2x}-e^{-6x}).}\ 
}
Defining $e^{2h}=e^{-2f} a^2$ and $e^{2\a }=e^{2 f} a^3= e^{5f +3 h}$,
we arrive at the following effective Lagrangian
\eqn\eff{
{\cal L}_{eff}=e^{2\alpha}(T-V)\ ,
}
where 
\eqn\cin{\eqalign{
T&=4{\dot \alpha}^2-3{\dot h}^2-3{\dot x}^2\ ,
\cr
V&={3\over 2g^2} e^{-4h-2x}- 6e^{-2h}-{1\over 2}
g^2(3e^{2x}+6e^{-2x}-e^{-6x})\ ,
}}
where ${\dot \alpha}\equiv {d\alpha\over dr}$, etc.

This effective Lagrangian together with the constraint $H=T+V=0$ 
arising
from diffeomorphism invariance in the radial coordinate determine the
second order equations of motion for $\a (r) ,\ h(r) ,\ x(r)$.
For supersymmetric configurations it is always possible to 
find a set of first order differential equations
whose solutions solve the equations of motion.
This can be done in the present case.
The Hamilton-Jacobi linear equations are 
$p_i={\p F\over \p \varphi _i}$, where $\varphi _i= \{ \a , h,x \}
$. Taking as principal function $F=e^{2\a}W(h,x)$ and if the potential
of the effective quantum mechanics  
can be written in terms of a superpotential $W$ as follows:
\eqn\sup{
V={1\over 12}(\partial_h W)^2+{1\over 12}(\partial_x W)^2-{1\over 
4}W^2\ ,
}
then one can find first order equations which solve the equations of 
motion.
The solution for \sup\ is
\eqn\supansatz{
W=3g e^x+ g e^{-3x} + {3\over  g} e^{-2h-x}\ .
} 
Thus  the  first-order (BPS) equations are:
\def\a{\alpha }
\eqn\aaq{
\dot \a ={1\over 4} W={3g\over 4} e^x+ {g\over 4} e^{-3x} +{3\over 4
g} e^{-2h-x}\ , 
}
\eqn\bbq{
\hskip-60pt\dot h=-{1\over 6}\partial_h W={1\over g} e^{-2h-x}\ ,
}
\eqn\ccq{\hskip+20pt\dot x=-{1\over 6}\partial_x W=-{g\over 2} e^x 
+{g\over 2} e^{-3x}+{1\over 2g} e^{-2h-x}\ .
}
By a change of radial coordinate 
$r\to \rho (r)={g^2\over 2}e^{2h(r)}$,  
and defining $u=e^{2x}$,  eqs. \aaq - \ccq\ take the form
  \eqn\ffq{
          {d\a \over d\rho }={1\over 4}(3 u +u^{-1}) +{3\over 8 \rho }\ 
,\ \ \ 
\ \ u^{1/2}(\rho)\ d\rho=g dr\ ,
          }
\eqn\eeq{
{du\over d\rho }- {1\over 2\rho}\ u =   1-u^2\ .
}
The general solution to eq. \eeq\ is given by
\eqn\sssq{
u(\rho )=e^{2x(\rho )}={c_0  I_{1/4}(\rho )  +(1-c_0)   I_{-1/4}( \rho 
)\over
    c_0  I_{-3/4}(\rho )  +   (1-c_0) I_{3/4}(\rho ) }\ ,
}
where we are using the standard notation $I_\nu (z)$ for the modified
 Bessel functions and $c_0$ is a parameter of the solution.
Then $\a (\rho )$   is determined by direct integration
of \ffq . These functions have a simple behavior, which will be 
described below.

\newsec{Ten-dimensional solution }

Since we have fully solved the $SO(4)$ gauged supergravity equations
one can write the corresponding Type IIB supergravity solution in ten
dimensions.  The expression of
the ten dimensional Einstein frame metric, the dilaton  
and the NS-NS three-form field strength are given 
in terms of the gauged
supergravity fields by
\eqn\tendim{\eqalign{
ds_{E}^2&=Y^{1/8}\left(\Delta^{1/4}ds_7^2+g^{-2}\Delta^{-3/4}T_{ij}^{-1}D\mu^i
D\mu^j\right)\cr
e^{2\phi}&={Y^{3/2}\over \Delta}\cr
H&={1\over 6g^2\Delta ^2} \epsilon_{i_1i_2i_3i_4}\bigg( U\
D\mu^{i_1}\wedge D\mu^{i_2}\wedge
D\mu^{i_3}\mu^{i_4}-3D\mu^{i_1}\wedge D\mu^{i_2}\wedge DT_{i_3j}
T_{i_4k}\mu^{j}\mu^{k} \cr
&-3g\Delta \ F^{i_1i_2}\wedge D\mu^{i_3} T_{i_4j}\mu^{j}\bigg),}}
where the $\mu^i$'s parametrise the transverse $\widetilde {\bf S}^3$
($\mu^i\mu^i=1$) and  
\eqn\moredef{\eqalign{
\Delta&=T_{ij}\mu^i\mu^j\ ,\cr
D\mu^i&=d\mu^i+gA^{ij}\mu^j\ ,\cr
U&=2T_{ik}T_{jk}\mu^i\mu^j-\Delta T_{ii}\ ,\cr
DT_{ij}&=dT_{ij}+g A^{ik}T_{kj}+gA^{jk}T_{ik}\ .
}}
The metric in the string frame is therefore
\eqn\stringframe{
ds_{st}^2=Y^{1/2}\left(ds_7^2+g^{-2}\Delta^{-1}T_{ij}^{-1}D\mu^i
D\mu^j\right).
}
Using \scal, \metr , we note that the string metric  has
no warped factor in the ${\bf R}^{1,2}$ part, 
with the identification we made earlier $y(r)=-4f(r)$.

Let us now write explicitly this new Type IIB supergravity solution.
First, 
we note that the seven dimensional gauge coupling is related to the 
brane
charge $N$ via the relation $1/g^2=N$. 
Using ~\stringframe , \metr , \scal\  the metric in the string frame 
and
the dilaton are given by ($\a'=1 $)  
\eqn\ddd{
ds^2_{st}=ds^2({\bf R}^{1,2})+ N\bigg[ e^{2 x}d\rho ^2+
{\rho\over 2} w_a w_a +{1\over  G}
\bigg( e^{-2x}\big( D\mu_1^2+D\mu_2^2+D\mu_3^2\big)+ e^{2x}
d\mu_4^2\bigg)\bigg] 
}
\eqn\ttd{
\hskip-280pte^{2\phi }=e^{2\phi_0} \rho e^{-2\a -x} G^{-1},}
where
\eqn\defineG{G\equiv  e^{-x-y/4}\Delta=\sin^2\psi +e^{-4x}\cos^2\psi \
,}
and 
 $\phi_0$ arises as the integration constant in 
eq.~\ffq\ for $\a $.
This is in the parametrization of $\widetilde {\bf S}^3$ given by
$$
\mu_1=\sin\psi\cos\phi_1 ,\ \  \mu_2=\sin\psi\sin\phi_1\cos\phi_2 ,\
\  \mu_3=\sin\psi\sin\phi_1\sin\phi_2 ,\ \ \mu_4 =\cos\psi .
$$  
Taking into account the $SU(2)_D$ truncation \gaugefi , one has the
following covariant derivatives
$$
D\mu_1= d\mu_1 - \ha w_3\mu_2+ \ha w_2\mu_3 ,\ \ 
D\mu_2= d\mu_2 + \ha w_3\mu_1- \ha w_1\mu_3 ,\ \
D\mu_3= d\mu_3 - \ha w_2\mu_1+ \ha w_1\mu_2 .\ \
$$
where we have used the gauge fields in the ansatz \gauge\ that
implement the required topological twisting. One can also write down
the three-form flux 
using \tendim\ , the curvature $F_{ij}$ given in  \curv\ and
$$
 dT_{ij}={dT_{ij}\over d\rho } d\rho\ ,
\ \ \ \ T_{ij}=e^{y/4} {\rm diag}\big( e^x,e^x
,e^x  ,e^{-3x} \big)\ ,
$$ 
$$
U=\ha e^{-6x+{y\over 2}}\bigg(1-4e^{4x}-e^{8x}+(e^{4x}-1)^2\cos 2\psi 
\bigg)\ ,
 \ \ \ y={4\over 5}(3h-2\a ), \ \ e^{2h}=2N\rho\ .
$$

\ifig\fone{The function $e^{-2x(\rho )}$ vanishes at some $\rho_0>0$ 
for
$c_0<0$. The figure shows the different curves for the values 
$c_0=-6,\ -0.5,\ 0,\ 0.4,\ 1.2$.}
{\epsfxsize=6cm \epsfysize=5.2cm \epsfbox{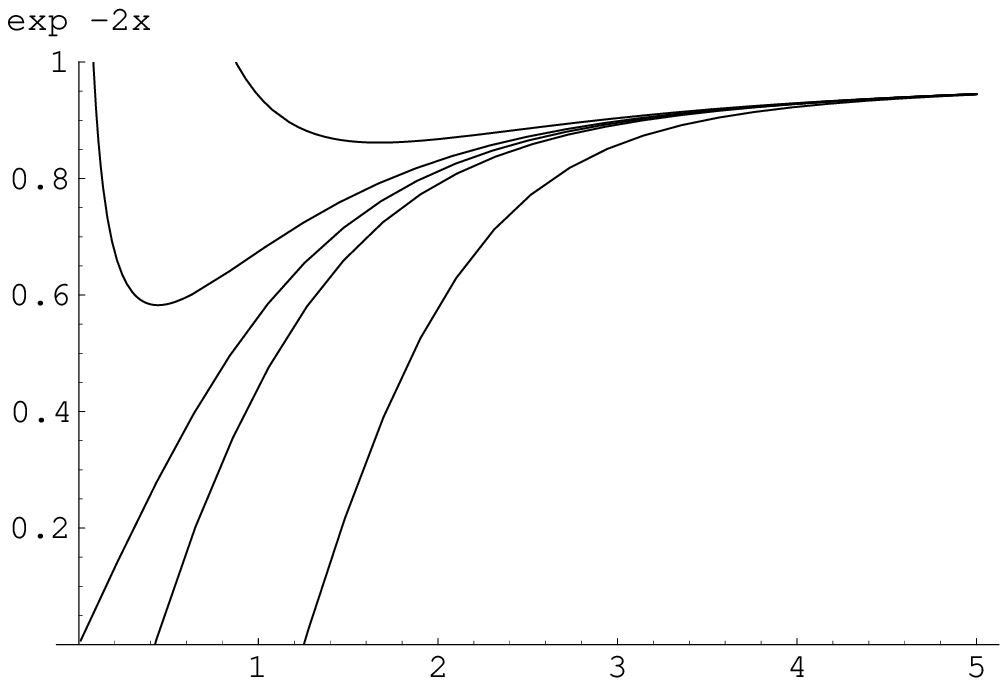}}

Let us now examine the behavior of the solution as a function of the
$c_0$ integration constant (see plot of $e^{-2x}$ in fig.1). There are
three different regimes: 
\medskip  
\noindent ${\bf a)}$ For $c_0<0$, the function $e^{-2x}=u^{-1}$ becomes
negative at $\rho<\rho_0$.  
Therefore, the space terminates at $\rho=\rho_0$ (since the scalar 
field $x$
is real). 
The value of $\rho_0 $ depends on $c_0$ and it decreases towards  zero 
as
$c_0\to 0^-$.
%
%
The behavior near $\rho = \rho_0$ is as follows:
\foot{For $c_0=0$ we have near $\rho =0$ that  
$
e^{-2x}\cong {2\over 3}\ \rho  \ , \ \  e^{2\a }\cong \rho^3 \ ,\ \  
e^{2\phi }\sim \rho^{-3/2}\ 
$
which results in a singular solution, which is an acceptable one
according to 
the criterium proposed by \malnun .}
$$
e^{-2x}=\td \rho +O\big(\td \rho ^2\big)\ ,\ \  \ \a \cong {3\over 
4}\log
\td\rho \ ,
$$
where $\td\rho =\rho-\rho_0$. The metric and dilaton near the 
singularity are then
given by
\eqn\nearh{\eqalign{
ds^2 ({\rm NS5})&\cong ds^2({\bf R}^{1,2})+{N\rho_0\over 2} w_aw_a+  
{N\over \td \rho }\bigg(d\td\rho ^2+ \td\rho^2 \big( d\phi_1^2 
+\sin^2\phi_1d\phi_2^2\big)+d\psi^2
\bigg)\ ,\cr
 e^{-2\phi_{\rm NS5} }&\cong e^{-2\phi_0}\sin^2(\psi){{\tilde 
\rho}\over \rho_0}.
}} 
We note that in this infrared limit the transverse ${\widetilde{\bf 
S}}^3$
degenerates to 
an ${\widetilde{\bf S}}^1\times {\widetilde{\bf S}}^2$.
The metric has a curvature singularity at $\rho=\rho_0 $.
There the dilaton $e^\phi $ goes to $\infty $.
The applicability of holography in the presence of singularities was
discussed in \malnun,\gubser. 
According to the criterium of \malnun , the singularity is a good one, 
since $g_{00}^E=-e^{-\phi }$ does not increase as we approach the 
singularity.
Here $g_{00}^E\cong  - (\rho- \rho_0)^{1/2}\to 0$ as
$\rho \to \rho_0 $. In the next section we will see that this
singularity is due to a brane source and it corresponds to a Coulomb
branch configuration of the dual three-dimensional ${\cal N}=2$ SYM 
theory.

\medskip
\noindent {\bf b)} For $0< c_0\leq 1$, $e^{-2x}$ goes to $+\infty$  as
$\rho \to 0$. 
For a generic value of the integration constant $1>c_0>0$,
 we have near $\rho=0$: 
\foot{If $c_0=1$,  the behavior near $\rho=0$ is
$
e^{-2x}\cong  (2\rho)^{-1} ,\    e^{2\phi} \sim \rho^{3/2} ,\ e^{2\a 
}\sim
\rho.$ 
The metric with this  singularity at $\rho=0$ is deemed to be 
unacceptable as a
dual description of a field theory \malnun .} 
\eqn\behav{
e^{-2x}\sim  \rho ^{-1/2}\ ,\ \ \ \  e^{2\phi} \sim \rho \ ,\ \ \ \  
e^{2\a }\sim
\rho^{3/4}\ .
}
We see that the metric has a
bad singularity, with $g_{00}^E$ diverging at $\rho=0$ and we discard
it as a possible gravity description of the infrared dynamics of a
gauge theory.
\medskip

\noindent {\bf c)} 
For $c_0>1$, the factor $e^{-2x}$ blows up at some $\rho_1 $,
$e^{2x}= (\rho -\rho_1)+O\big((\rho-\rho_1)^2\big)$ and
$e^{2\phi}={\rm const.}(\rho-\rho_1)$. Thus  $e^{-\phi }\to \infty $,
so the singularity is again a bad one and we will discard these 
solutions.
\smallskip


In some respects, the structure of the solution resembles  the
solution found in \refs{\jerome , \zaff}, and the solution of \PILCH ,  describing $\N=2 $ 3+1 SYM. The 
parameter
$c_0$ is the counterpart of the parameters $k$ and $\gamma $ in 
\jerome , \PILCH , \bupepol .

 \medskip

The UV behavior near $\rho =\infty $ is, for any $c_0$, given by
 $u=e^{2x}\cong 1+{1\over 4 \rho }$ (i.e. $x\to 0$), $\phi \sim -\rho\
 ,\ \a = \rho +\ha \log\rho\ $. This is in fact the same leading
 behavior as the solution 
in \mnas\ (but here there is no subleading term of the form $\log\rho $ 
in $\phi $) . The metric is
\eqn\dddd{
ds^2_{st}=ds^2({\bf R}^{1,2})+ N\bigg( {1\over 4} d\rho ^2+{\rho \over 
2} w_a w_a +
 D\mu_1^2+D\mu_2^2+D\mu_3^2+  d\mu_4^2\bigg)
}
This describes the near horizon limit of the NS5 brane wrapped on 
3-sphere, with a twist.

\medskip
In conclusion, we  expect that the present supergravity solution with
$c_0<0$  is dual to three dimensional $\N =2$ $SU(N)$ SYM.
Below we discuss this correspondence in more detail.

\newsec{Field theory from the supergravity description}

We will now show that the present $\N =2 $ model
based on the one-parameter family of solutions
with $c_0<0$ is  a gravitational  dual of  the Coulomb branch of $\N 
=2$ SYM
theory in three space-time dimensions.

To properly describe the infrared dynamics of the gauge theory on the
NS5-branes one has to perform an S-duality transformation and consider
D5-branes \malis . The corresponding D5-brane solution is
given by
\eqn\ddd{
ds^2({\rm D5})=e^{-\phi_{\rm NS}} ds^2({\rm NS5})\ ,\ \ \
 \phi_{\rm D5}=-\phi_{\rm NS}
}
$$
C^R_{(2)}=-B_{(2)}^{NS}\ ,\ \ \ \ \ C^R_{(6)}=-B^{NS}_{(6)}
$$
where $B^{NS}_{(6)}$ is dual to  $B_{(2)}^{NS}$ and $C^R_{(6)}$ is
dual to $C^R_{(2)}$
\eqn\hodged{
dB^{NS}_{(6)}= e^{-2\phi_{\rm NS5}}*H^{NS}.}
Since there is no flux of $H^R$ through the ${\bf S}^3$ which the
D5-branes wrap, there is no induced Chern-Simons term  and the
supergravity solution describes
the large $N$ limit of
$D=2+1$ $SU(N)$ SYM theory with no Chern-Simons term.
This field theory is classically related to the dimensional
reduction of $\N=1$  
SYM in four dimensions.
Perturbatively this theory has a moduli space of vacua parametrized by
the scalars in the vector multiplet together with the dual photons.
An interesting property of this gauge theory is that BPS monopole
configurations induce a non-perturbative superpotential which lifts
the Coulomb branch \ahwit . In the 't Hooft limit these
effects are of order $e^{-N}$ and are not seen in the supergravity
approximation. The supergravity analysis exhibits the existence, in
perturbation theory, of a Coulomb branch of vacua.


The fact that the supergravity solution behaves as $g_{00}^E\cong - 
(\rho-
\rho_0)^{1/2}\to 0$ as 
$\rho \to \rho_0 $, 
means that, in this region, excitations in the background of fixed
proper energy correspond  
to very small energy excitations in the dual field theory.
Therefore, despite the metric being singular at $\rho= \rho_0 $, the
singularity type can capture information about infrared physics of the
gauge theory \malnun.

The background exhibits also an interesting property. In the UV
region, we have seen that  
it reduces essentially to the near horizon limit of NS5-branes
wrapped on ${\bf S}^3$, with a twist. 
In the IR region, the metric and dilaton are  given by eq.~\nearh . 
It describes NS5-branes with a flux on ${\widetilde{\bf S}}^2\times
{\widetilde{\bf S}}^1$, 
i.e. smeared  
on a ring parametrized by $\psi $, and wrapped on ${\bf S}^3$.
A similar feature was observed in the solution of \refs{\jerome,\zaff}.

In order to study the moduli space, we follow \refs{\jpp,\bupepol,\ejp } and consider
the effective action 
of a probe D5 brane, which corresponds to breaking $SU(N)\rightarrow
SU(N-1)\times U(1)$ (see also \khawar , \jonlopageb \ for other probe
calculations  with four
supercharges).
In the present D5 brane solution, the NS two-form vanishes,
and the effective action is given by
\eqn\prob{
S=- {1\over (2\pi)^5{\a'}^3 } \int \bigg(d^6y \ e^{-\phi_{\rm 
D5}}\sqrt{-\det\big[G_{ab}+2\pi\a' F_{ab}\big] }
-\  C_{(6)}\bigg)\ ,
}
where $F_{ab}$ is the world-volume abelian gauge field. 
We fix the static gauge where the world-volume coordinates are 
identified
with $x_{0,1,2}, \theta ,\beta ,\varphi $, so that the four scalar 
fields 
$\rho ,\psi ,\phi_1,\phi_2$ are functions of these coordinates.
The effective action contains kinetic terms and a potential for these
scalar fields.
The potential can be determined by setting to constants the scalar 
fields
$\rho ,\psi ,\phi_1,\phi_2$ and $F_{ab}=0$.
The locus where the potential vanishes determines the moduli space.
For the present background, the potential has a complicated form,
but it is easy to see that it vanishes at
$\rho=\rho_0$, $c_0 < 0$. The reason is as follows.
For $\rho\cong \rho_0$, the metric and dilaton take the simple form as 
in eq.~\nearh , 
where the five brane is smeared on ${\bf S}^1$ parametrized by $\psi $. 
We find 
$$
e^{-\phi_{\rm D5}}\sqrt{-\det\big[G_{ab}
]}=
e^{-2\phi_0}({N\over 2})^{3/2}\rho_0^{1/2}\td\rho\sin\theta\sin^2\psi\ 
.   
$$  
Using
$$
 U\cong - e^{2x+y/2} \sin^2\psi\ ,\ \ \ \Delta \cong
 e^{x+y/4}\sin^2\psi\ ,
$$
one can see that in this region $\rho\cong\rho_0$ the three-form field 
strength reduces to
\eqn\hns{
H^{NS}=- N \sin\phi_1 \ d\phi_1\wedge d\phi_2 \wedge d\psi
}
As expected, this is the 3-form corresponding to the five brane
\nearh\ with flux 
on $\widetilde {\bf S}^2\times \widetilde {\bf S}^1$.
The dual 6-form is then
\eqn\huu{
C_{(6)}^R= e^{-2\phi_0}  ({N\over 2})^{3/2} \rho_0^{1/2}\td\rho
 \sin\theta \sin^2\psi 
 \ dx^0\wedge dx^1\wedge dx^2\wedge d\varphi \wedge d\theta \wedge
 d\beta \ .
}
Therefore
$$
S_{\rm pot}(\rho=\rho_0)=0\ ,\ \ \ {\rm for}\ c_0<0\ .
$$
Thus the moduli space is at $\rho=\rho_0$, $c_0<0$.

The dimension of the moduli space is determined by the number of 
massless scalar
fluctuations at $S_{\rm pot }=0$. Thus we are to look at the
Born-Infeld lagrangian 
for the solutions with $c_0<0$ in the limit $\rho\to \rho_0$.
We get the following kinetic terms for the bosonic fields
\eqn\kkin{
S_{\rm kin}=
- \int d^3x\  
\big( L^2 \sin^2\psi \partial_i\psi \partial_i\psi
+  4 g_3^2  \partial_i A \partial_i A \big)\ ,\
 }
$$
L^2= {N e^{-2\phi_0}\over  8 g_3^2\pi ^2\rho_0}\ ,\ \ \ \ 
g_3^2={(2N\rho_0)^{3/2}\over 4\pi }\ ,
$$
where we have dualized the gauge field into a compact scalar $A$ of 
period
$2\pi $.
It is is easy to see that there is no mass term for $\psi $ at 
the locus of $S_{\rm pot}=0$, so $\psi $ and $A$ are  moduli.
Thus we find a two-dimensional moduli space, as expected for an $\N
=2$ the gauge theory in three space-time dimensions, which gives
further evidence of the supersymmetry of the solution.
Defining $\eta =\cos\psi $, the moduli space metric takes the form
\eqn\modul{
ds^2_{\rm Mod}=L^2 d\eta ^2 +4g_3^2 dA^2\ ,\ \ \ \ |\eta|\leq 1\ ,\ 
}
which describes a flat closed strip. It is a complex moduli space with 
a
Kahler metric as 
expected by supersymmetry. \foot{A Kahler metric for four dimensional 
$\N=4 $
broken to $\N =1$ SYM via mass terms
was found in \jonlopage .}
\smallskip

In conclusion, 
the supergravity solution captures the  expected features of the large
$N$ limit of three dimensional $\N =2$ SYM theory in the Coulomb 
branch.

\medskip\bigskip

\noindent{\bf Acknowledgements}
\medskip
We would like to thank K. Pilch  and N. Warner for discussions, and
CERN for hospitality during the course of this work.
J.R. would also like to thank Caltech for hospitality. The work of
J.G. is supported in part by the National Science Foundation under
grant No. PHY99-07949 and by the DOE under grant No. DE-FG03-92ER40701.
The research of J.R.
is supported by Universidad de Buenos Aires 
and Conicet.

\vfill\eject

\appendix{A}{NS5 branes with $SU(2)_D$ gauge fields for 3+1 SYM}

As explained in sect. 2, the natural $SU(2)$ truncation that can lead 
to a non-abelian solution with eight supercharges is 
 when one truncates the $SO(4)$ gauge
fields to $SU(2)_D=(SU(2)_L\times SU(2)_R)_{diag}$.
Supersymmetry alone fixes the required $SU(2)$ truncation to be the
$SU(2)_D$ one. 

Having a solution of wrapped five branes describing
$2+1$ super Yang-Mills with $\N=2$ supersymmetry, it is natural to look
for a solution with $SU(2)_D$ gauge fields
dual to $\N=2 $ $3+1$ super Yang-Mills.
In this appendix we write down the ansatz for a collection of wrapped
NS5-branes on the ${\bf S}^2$ of the Eguchi-Hanson geometry. The
seven dimensional metric  has the following form
\eqn\metr{
ds_7^2=e^{2f(r)}(ds^2({\bf 
R}^{1,3})+dr^2)+a(r)^2(d\theta^2+\sin^2\theta
d\phi^2)\ .}
The non-abelian ansatz we take for the gauge fields is
\eqn\gaugeans{\eqalign{
gA^{1}&=A(r)d\theta\ ,\cr
gA^{2}&=-A(r)\sin\theta d\phi\ ,\cr 
gA^{3}&= \cos\theta d\phi\ ,
}}
which has the following field strengths
\eqn\fieldsas{\eqalign{
gF^{1}&= {\dot A(r)} dr\wedge d\theta\ ,\cr
gF^{2}&=-{\dot A(r)} \sin\theta dr\wedge d\phi\ ,\cr 
gF^{3}&= \sin\theta (A(r)^2-1) d\theta \wedge d\phi\ ,
}}
where ${\dot A}={dA \over dr}$.
These gauge fields preserve the $SU(2)$ symmetry of the ${\bf S}^2$
and have the property that they are pure gauge when $A=1$.
Note that the gauge fields satisfy the condition \consistent\ 
$F_a\wedge *F_b=0$,
$a\neq b $. 
Lastly, we will take the two scalars to depend on the radial
coordinate $r$, so we also   have $x(r)$ and $y(r)$.

One way of solving the equations of motion is to insert the ansatz
into the Lagrangian \truncLagr\ and solve the equations of motion of
the effective quantum mechanical model with $r$ being interpreted as
time. One must also impose the zero energy condition on the
system. Plugging into the Lagrangian our ansatz one gets the following
contributions
\eqn\pluffedin{\eqalign{
{\cal L}_R&=e^{3f}a^2\left(2{e^{2f }\over a^2}+2{\left({\dot
a}\over a\right)^2}+16{{\dot
a}\over a}{\dot f}+12 {\dot f}^2\right)\cr
{\cal L}_{scal}&=-e^{3f}a^2\left({5\over 16}{\dot y}^2+3{\dot
x}^2\right)\cr
{\cal L}_{gauge}&=-{1\over 2g^2}e^{3f}a^2 e^{-2x-y/2}\left( 2{{\dot
A}^2\over a^2}+{(A^2-1)^2\over a^4}e^{2f}\right)\cr
{\cal L}_{pot}&={1\over 2}e^{5f}a^2 g^2 
e^{y/2}(3e^{2x}+6e^{-2x}-e^{-6x})}.
}
The ten dimensional  string frame metric is of the form 
$ds^2_{st} = e^{2f+y/2} ds^2 ({\bf R}^{1,3})+...$\ Since we are interested in 
NS five brane solutions, it is natural (as in the solution of sect. 3) to look for
solutions with no warp factor in the parallel directions, i.e. $y=-4f$.
This is consistent with the equations of motion. Indeed, consider the equation
of motion for $f$ following from \pluffedin . It contains  a term having second derivatives of $a$. 
Using the equation for $a$, one can express that term in terms of first derivatives.
The resulting equation is the same as the equation for $y$, after setting $y=-4f$ and using the Hamiltonian
constraint
$$
0=H=a^2 e^{3f}\bigg( 2{\dot a^2\over a^2} +16  {\dot a\over a}\dot f+12  \dot f^2 -{1\over g^2a^2} e^{-2x-y/2} \dot A^2
-3  \dot x^2-{5\over 16} \dot y^2 \bigg)
$$
$$
 -\ 2e^{5f} + {1\over 2g^2} e^{5f-2x-y/2} {(A^2-1)^2\over a^2}  -{1\over 2} a^2 g^2 
e^{5f+y/2}(3e^{2x}+6e^{-2x}-e^{-6x})\ .
$$
Setting $y=-4f $, 
and  defining 
\eqn\funcdef{\eqalign{
e^{2\alpha}=e^{3f}a^2\cr
e^{2h}=e^{-2f}a^2,}}
one obtains the following Lagrangian
\eqn\effef{
{\cal L}_{eff}=e^{2\alpha}(T-V),}
with
\eqn\tform{
T=4{\dot \alpha}^2-2{\dot h}^2-3{\dot x}^2-{1\over
g^2}e^{-2h-2x}{\dot A}^2}
and
\eqn\potform{
V=-2e^{-2h}+{1\over 2g^2}e^{-4h-2x}(A^2-1)^2-{1\over
2}g^2(3e^{2x}+6e^{-2x}-e^{-6x}).
}
It would be interesting to search for BPS equations using this
effective Lagrangian.

\listrefs

\bye